\documentclass[aps,prb,twocolumn,groupedaddress,showpacs]{revtex4}
\usepackage{graphicx}
\usepackage{dcolumn}
\usepackage{bm}
\usepackage{amsmath,amsfonts,amssymb,amsthm,verbatim}
\usepackage[ansinew]{inputenc}
\usepackage{color}

\usepackage{epsfig}
\usepackage{wasysym}

\begin{document}

\title{Scaling behavior in a multicritical one-dimensional topological insulator}

\author{M. Malard$^{1}$, H. Johannesson$^{2}$, W. Chen$^{3}$}
\affiliation{$\mbox{}^1$Faculty of Planaltina, University of Brasilia, 70904-910, Federal District, Brazil}
\affiliation{$\mbox{}^2$Department of Physics, University of Gothenburg, SE 412 96, Gothenburg, Sweden}
\affiliation{$\mbox{}^3$Department of Physics, PUC-Rio, 22451-900, Rio de Janeiro, Brazil}

\begin{abstract}

A class of Aubry-Andr\' e-Harper models of spin-orbit coupled electrons exhibits a topological phase diagram where two regions belonging to the same phase are split up by a multicritical point. The critical lines which meet at this point each defines a topological quantum phase transition with a second-order nonanalyticity of the ground-state energy, accompanied by a linear closing of the spectral gap with respect to the control parameter; except at the multicritical point which supports fourth-order transitions with parabolic gap-closing. Here both types of criticality are characterized through a scaling analysis of the curvature function defined from the topological invariant of the model. We extract the critical exponents of the diverging curvature function at the non-high symmetry points in the Brillouin zone where the gap closes, and also apply a renormalization group approach to the flattening curvature function at high symmetry points. We also derive a basis-independent correlation function between Wannier states  to characterize the transition. Intriguingly, we find that the critical exponents and scaling law defined with respect to the spectral gap remain the same regardless of the order of the transition.

\end{abstract}


\maketitle

\section{Introduction}

In symmetry-protected gapped topological materials, each topological phase of the material is characterized by a integer-valued topological invariant calculated according to the symmetry and dimension of the system\cite{Schnyder08,Kitaev09,Ryu10,Chiu2016}. A change of the topological invariant, always accompanied by the vanishing of the gap in the energy spectrum of the corresponding system, indicates the occurrence of a topological quantum phase transition (TQPT). However, since the topological invariant remains constant as the system approaches a TQPT, it does not provide information on the critical behavior of observables near the transition.

Starting to address this issue, a new branch in the study of TQPTs is currently being developed. Recent works \cite{Continentino1,Continentino2,RoyGoswamiSau,Chen1,Chen2,Chen3,Griffith,Niewenburg,Chen4,Rufo,Continentino3,Chen19_book_chapter,Molignini20} have explored the question as to whether TQPTs possess a critical behavior analogous to that of symmetry-breaking phase transitions, with scaling of observables controlled by universal critical exponents \cite{Sachdev,ContinentinoBook}. One approach\cite{Chen1,Chen2} suggests that the topological invariants of the Altland-Zirnbauer classification \cite{Chiu2016} can be expressed as an integral of a function of momentum and the Hamiltonian parameters - a \emph{curvature function} - whose asymptotic scaling behavior near the transition is governed by critical exponents and can be analyzed using a curvature renormalization group (CRG) method. Moreover, the relation between the curvature function and the overlap between the bulk Wannier states of the system (weighted by an operator determined by the relevant symmetry class of the Altland-Zirnbauer classification) yields a correlation length which diverges at criticality \cite{Chen4}. In another proposal, the scaling laws of a TQPT follow from the behavior of the localization length of the topological edge states \cite{Continentino2}, which was later shown to coincide with the Wannier state correlation length\cite{Chen3,Chen4}.

Recent investigations of TQPTs have uncovered that a number of topological systems exhibit the fascinating phenomenon of quantum multicriticality. As it is known, when two or more critical lines intersect at a point $-$ a {\em multicritical point} $-$ a sudden change of an otherwise smooth critical behavior may occur \cite{ChaikinLubensky}. The study of various aspects of multicriticality $-$ scaling functions, borderline dimensions, critical and crossover exponents, amplitude ratios, and other properties $-$ has spawned a huge literature within the theory of classical phase transitions \cite{FisherBook}, producing insights and results that have informed our understanding also of multicritical quantum phase transitions \cite{CarrBook}. Drawing relevance to topologically ordered matter\cite{Wen}, the importance of multicriticality was recognized early on, and there is now a growing body of works on the subject, starting with Refs. \onlinecite{Xu} and  \onlinecite{Tupitsyn}. Examples of multicritical points in the phase diagrams of gapped symmetry-protected topological matter include the Haldane model for a Chern insulator \cite{Haldane}, the Creutz model with induced superconductivity \cite{Sticlet}, and the dimerized Kitaev chain \cite{Wakatsuki}. The CRG approach to topological multicriticality was applied in Ref. \onlinecite{Molignini20} to a periodically driven Floquet-Chern insulator, showing that the coexistence of a linear Dirac-like transition with a quadratic nodal looplike transition implies multiple universality classes and scaling laws. Another take on topological multicriticality was presented by Rufo {\em et al.} \cite{Rufo} who identified a multicritical line with unusual scaling behavior in the Su-Shrieffer-Heeger model with an added synthetic potential.

In the present work we draw on the approach introduced in Refs. \onlinecite{Chen3,Chen4} to investigate the phase transitions uncovered in Ref. \onlinecite{Malard20} where a generalized Aubry-Andr\'e-Harper model \cite{AubryAndre,Harper} for spin-orbit coupled electrons on a one-dimensional (1D) lattice was shown to display a checkerboard-like phase diagram supporting trivial and topological gapped phases separated by critical lines. Crossing a single critical line gives a TQPT characterized by a jump of the topological invariant, accompanied by a second-order nonanalyticity of the ground state energy and a linear closing of the spectral gap with respect to the control parameter. At multicritical points $-$ defined by the intersections between two critical lines $-$ the ground-state energy was still found to be nonanalytical, although only at fourth order, and the gap was still found to close, although parabolically, even though these points may be crossed without a jump of the topological invariant or change of symmetry.

In this article a scaling analysis of the curvature function for both usual TQPTs and the multicritical points unveils a remarkable property: Despite the orders of the transitions and of the gap-closings being different for the two types of transitions, the scaling of the curvature with respect to the gap is governed by the same critical exponents and scaling law. This leads us to suggest that these critical exponents and scaling law are universal and can be generalized to higher order and more complicated TQPTs, provided the curvature function evolves continuously with the Hamiltonian parameters.

The article is laid out as follows: In Sec. II we present the representative Aubry-Andr\'e-Harper-type model which we shall study, discuss its relevant features and possible symmetry classes. In Sec. III we revisit the phase diagram of the model obtained in Ref. \onlinecite{Malard20}, and the main properties of the gapless spectra associated to its critical lines. The curvature function is derived and its divergence at non-high symmetry points in the Brillouin zone (BZ) is analyzed in Sec. IV, from which we extract the critical exponents and scaling law. In Sec. V we derive a relation between the curvature function and a ``skew polarization" \cite{Mondragon}, the Fourier transform of which yields a ``skew correlation function" between bulk Wannier states characterized by a diverging correlation length. In Sec. VI, a CRG analysis is carried out around a high-symmetry point in the BZ which is a faster method (in computer time) for obtaining the phase diagram than the approach used in Ref. \onlinecite{Malard20}, and yields additional information regarding the stability of the critical lines and multicritical points. Our conclusions and final remarks are presented in Sec. VII.

\section{Model}

We consider a one-dimensional (1D) lattice with $N$ sites populated by electrons with nearest-neighbor hopping, two types of spin-orbit interactions and a spatially modulated contribution to each of those terms. The lattice tight-binding Hamiltonian writes:
\begin{equation}\label{H}
H\,=\,\sum^{N}_{n=1}\sum_{\alpha,\alpha'}\,h_{\alpha\alpha'}(n)\,c^{\dag}_{n,\alpha}\,c_{n+1,\alpha'}\,+\,\mbox{H.c.},
\end{equation}
where
\begin{equation}\label{h}
h_{\alpha\alpha'}(n)\,=\,t(n)\delta_{\alpha\alpha'}+i\gamma_{\text{R}}(n)\sigma_{\alpha\alpha'}^{y}+i\gamma_{\text{D}}(n)\sigma_{\alpha\alpha'}^{x},
\end{equation}
and $c^{\dag}_{n,\alpha}$ ($c_{n,\alpha}$) is the creation (annihilation) operator for an electron at site $n$ with spin projection ${\alpha}\!=\,\uparrow,\downarrow$ along a $z$-quantization axis, $\delta_{\alpha\alpha'}$ is the Kronecker delta and $\sigma^{x(y)}$ is the $x$ ($y$) Pauli matrix. With this choice of coordinates, the chain is along the $x$-axis. The spatially modulated parameters $t(n)$, $\gamma_{\text{R}}(n)$ and $\gamma_{\text{D}}(n)$ are modeled as $X(n)=-X-X'\cos(2\pi qn+\phi)$ with $X=t,\gamma_{\text{R}},\gamma_{\text{D}}$ ($X'=t',\gamma'_{R},\gamma'_{D}$) being, respectively, the strength of a uniform (modulated) hopping, Rashba spin-orbit interaction and Dresselhaus spin-orbit interaction; $a/q$ is the wave length of the modulation, with $a$ the lattice spacing and $1/q$ an integer number; and $\phi$ is a phase shift. This Hamiltonian belongs to the class of Aubry-Andr\'e-Harper models \cite{AubryAndre,Harper}, with the restriction of a commensurate periodicity between the external modulation and the underlying lattice. A model of this kind was investigated in Ref. [\onlinecite{Ganeshan}] for spinless particles. The spinful extension defined by Eqs. (\ref{H})-(\ref{h}) was studied in Refs. \onlinecite{Malard20} and \onlinecite{Malard18} for $q=1/4$, and it was found that the enlarged parameter space due to the presence of spin-orbit interactions produces a richer phase diagram than that of the spinless case.

To see how, let us follow the analysis of Ref. \onlinecite{Malard20} and impose periodic boundary conditions on the Hamiltonian $H$ in Eq. (\ref{H}).  $H$ is invariant under translations by a unit cell on a chain with $M=Nq$ cells and $r=1/q$ sites per cell. Performing a rotation of basis that diagonalizes the uniform part of $H$ in spin space, followed by a Fourier transform, yields the Bloch Hamiltonian represented by the $2r\times2r$ matrix
\begin{equation} \label{Hk}
{\cal H}(k)\,=\, \begin{bmatrix}
    0 & Q(k) \\
    Q^{\dagger}(k) & 0 \\
\end{bmatrix},
\end{equation}
with the $r\times r$ matrix $Q(k)$ given by
\begin{equation} \label{Qk}
Q(k)\,=\, \begin{bmatrix}
    A_{1} & 0 & 0 & \dots & 0 & A^{\ast}_{r}z\\
    A^{\ast}_{2} & A_{3} & 0 & \dots & 0 & 0 \\
    \vdots & \vdots & \vdots & \dots & \vdots & \vdots \\
    0 & 0 & 0 & \dots & A^{\ast}_{r-2} & A_{r-1} \\
\end{bmatrix},
\end{equation}
where $z=e^{-ik}$ and
\begin{equation} \label{A}
A_{n}\,=\, \begin{bmatrix}
    \alpha_{n}^{+} & \beta_{n}\\
    \beta_{n} & \alpha_{n}^{-}
\end{bmatrix}
\end{equation}
are $2\times 2$ matrices whose diagonal (off-diagonal) entries are given by spin-conserving (spin-flipping) hopping amplitudes $\alpha_{n}^{\tau}$ ($\beta_{n}$). In terms of the original parameters, these amplitudes read:
\begin{eqnarray}\label{alphabeta}
\nonumber \alpha_{n}^{\tau}&=&-[t+i\tau\gamma_{\text{eff}}]-\\
\nonumber &-&[t'+i\tau(\gamma_{RR}+\gamma_{DD})]\cos(2\pi qn+\phi),\\
\beta_{n}&=&i(\gamma_{RD}-\gamma_{DR})\cos(2\pi qn+\phi),
\end{eqnarray}
with $\gamma_{\text{eff}}=\sqrt{\gamma^{2}_{R}+\gamma^{2}_{D}}$, $\gamma_{RR}=\gamma'_{R}\gamma_{\text{R}}/\gamma_{\text{eff}}=\gamma'_{R}\cos\theta$, $\gamma_{DD}=\gamma'_{D}\gamma_{\text{D}}/\gamma_{\text{eff}}=\gamma'_{D}\sin\theta$, $\gamma_{RD}=\gamma'_{R}\gamma_{\text{D}}/\gamma_{\text{eff}}=\gamma'_{R}\sin\theta$, $\gamma_{DR}=\gamma'_{D}\gamma_{\text{R}}/\gamma_{\text{eff}}=\gamma'_{D}\cos\theta$, and $\tau=\pm$ labeling the spin projections along the new quantization axis. It follows that fixing the values of $t,t',\gamma'_{R},\gamma'_{D}$ and $q$, the model gets parametrized by the parameters $\gamma_{\text{eff}}$, $\theta$ and $\phi$. Details of the formalism can be found in the Supplemental Material to Ref. \onlinecite{Malard20}.

Together with its spinless variations, the class of models defined by Eqs. (\ref{Hk})-(\ref{alphabeta}) realize six out of the ten symmetry classes in the Altland-Zirnbauer classification\cite{Schnyder08,Kitaev09,Ryu10,Chiu2016}, with topologically nontrivial realizations only for even number of sites per unit cell. Indeed, for even $r$, the $2r\times2r$ spinful Bloch Hamiltonian ${\cal H}(k)$ is invariant under chiral ($S$) and time-reversal ($T$), and thus also particle-hole ($C$), symmetries \cite{Malard20} and hence - when enforcing all symmetries of the Hamiltonian on the allowed perturbations - belongs to class CII of the Altland-Zirnbauer classification. In their spinless version, the $2\times 2$ matrices $A_{n}$ become complex numbers, resulting in a violation of $T$ and, thus, placing the now $r\times r$ version of ${\cal H}(k)$ in class AIII. Imposing real hopping amplitudes restores $T$ in the previous case, thus changing the symmetry class to BDI. It follows that the gapped phases of both the spinful and spinless chains with an even $r$ are characterized by a $Z$-topological invariant (or $2Z$-invariant in case of CII). However, when $r$ is odd, the $2r\times2r$ spinful ${\cal H}(k)$ breaks $S$ and, thus, the symmetry class changes from CII to AII. As before, the spinless version additionally breaks $T$, sending the model to class A. Reinstating $T$ by imposing real hoppings changes the symmetry class to AI. Thus, an odd $r$ implies that both the spinful and spinless chains have only a topologically trivial gapped phase. This leads to the observation that the same system changes from trivial to topological simply by adding one site in the unit cell.

From the three topologically nontrivial possibilities above, the class-CII chain is the one of interest here. While the spinless chain with even $r$ has a one-dimensional topological phase diagram parametrized by $\phi$ only (for, in this case, only the kinetic hopping term is present), the phases of the spinful counterpart exist in the three-dimensional $\theta\times\phi\times\gamma_{\text{eff}}$ parameter space. This enlarged topological phase diagram opens up the possibility for multicriticality. As shown in Ref. \onlinecite{Malard20}, the choice of $r=4$ sites per unit cell provides the minimal realization of the model which supports a multicritical phase diagram. The particle-hole symmetry of class CII enforces that the band structure of any second-quantized model with Bloch Hamiltonian in this class is half filled, that is, referring to Eq. (\ref{Hk}), the lowest (highest) $r$ bands are completely filled (empty), yielding a band insulator with a gap about zero energy. The ground state of this band insulator - formed by the Slater determinant of the Bloch single-particle states of the filled bands - is unique in both the topologically trivial and nontrivial insulating phases. With $r=4$, the model acquires an off-centered mirror symmetry when $\phi=\pi/4$ and this symmetry (in addition to those of the CII class) forces the band gap to close at zero energy. In this way, multicritical lines (i.e. lines formed by a dense set of multicritical points) are generated from the crossings of the critical plane defined by $\phi=\pi/4$ and a set of critical surfaces supporting accidental band crossings.

\section{Phase diagram}

To explicitly show how multicriticality arises in our class CII model, we revisit the phase diagram obtained in Ref. \onlinecite{Malard20}. For that purpose, we recall that the topological invariant $W$ characterizing the gapped phases of a 1D system in symmetry class CII is a $2Z$-winding number \cite{Schnyder08,Kitaev09,Ryu10,Chiu2016} which, for a Bloch Hamiltonian cast in the form of Eq. (\ref{Hk}), is defined as the number of times that the function $\text{det}[Q]$ winds around the origin of the complex plane as $k$ runs through the BZ from $-\pi$ to $\pi$. Writing $\text{det}[Q]=Re^{i\delta}$, it follows from the definition that $W=-(2\pi)^{-1}\int d\delta$ (where the minus sign is introduced to make $W>0$ for $\text{det}[Q]$ winds clockwise, i.e. $d\delta<0$). Or, equivalently,
\begin{equation} \label{W}
W({\bf M})\,=\,-\frac{1}{2\pi}\int_{-\pi}^{\pi}\partial_{k}\,\delta({\bf M},k)\,\,dk,
\end{equation}
where ${\bf M}\equiv(\theta,\phi,\gamma_{\text{eff}})$ is a vector in the three-dimensional parameter space.

Using that $i\delta=\ln(\text{det}[Q])-\ln(R)$, Eq. (\ref{W}) can be rewritten as
\begin{equation} \label{Walternative}
W({\bf M})\,=\,-\frac{1}{2\pi i}\int_{-\pi}^{\pi}\frac{\partial_{k}\,\text{det}[Q({\bf M},k)]}{\text{det}[Q({\bf M},k)]}\,\,dk,
\end{equation}
where we used that, differently from $\text{det}[Q]$, $R$ is a single-valued real function of $k$ with $R(-\pi)=R(\pi)$ and thus the integral of $\partial_{k}\ln(R)$ over the BZ vanishes.
\begin{figure*}[htpb]
\includegraphics[width=18cm]{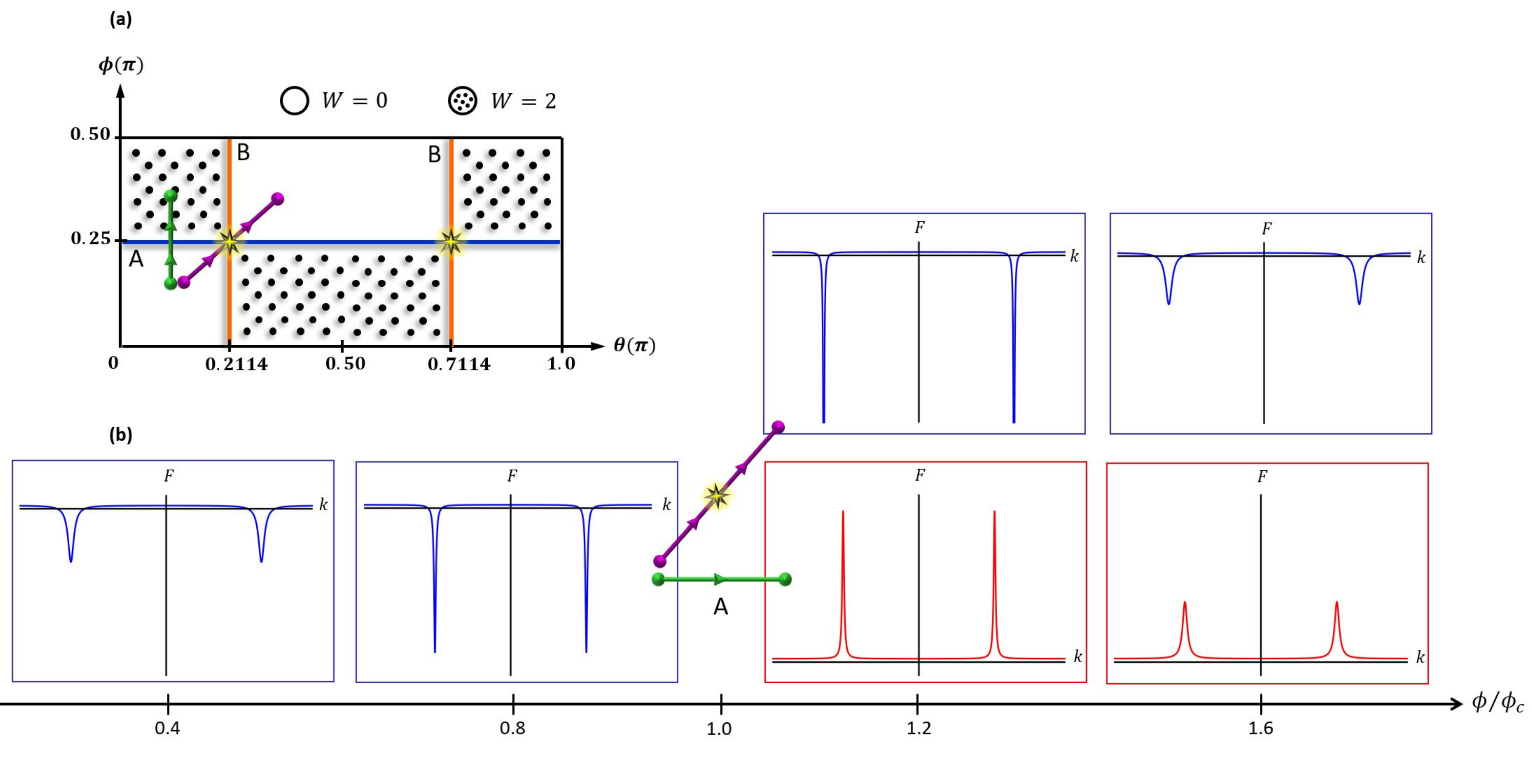}
\caption{(Color online) (a) Winding number $W$ on the $\theta\times\phi$ plane for $\gamma_{\text{eff}}=2.5$ (in arbitrary units). The phase diagram consists of topologically nontrivial gapped phases (dotted) where $W=2$ and trivial gapped phases (empty) where $W=0$, separated by critical lines $A$ and $B$ which cross at two multicritical points marked by yellow stars. The green (magenta) path represents a TQPT (transition between equal-$W$ regions) across critical line A (a multicritical point). (b) Curvature $F$ as a function of momentum $k$ for $\gamma_{\text{eff}}=2.5$ and $\theta=0.1\pi$, with $\phi$ varying along the green path shown in Fig. 1(a); $\theta$ and $\phi$ varying along the magenta path shown in Fig. 1(a). For both paths, when $\phi\neq\phi_{c}=\pi/4$, $F$ peaks around two points in the BZ which converge to the points where the gap closes when $\phi=\phi_{c}$, with progressively sharper peaks as $\phi\rightarrow\phi_{c}$. For the green (magenta) path the peaks flip (keep) their orientation at the transition.}
\end{figure*}

In Ref. \onlinecite{Malard20}, $W$ was numerically computed in the $\theta\times\phi\times\gamma_{\text{eff}}$ parameter space for $r=4$, yielding a phase diagram consisting of topologically nontrivial and trivial gapped phases separated by intersecting gapless (critical) surfaces, with the intersections defining multicritical lines. Fig. 1(a) shows a cross-section of that phase diagram for $\gamma_{\text{eff}}=2.5$ (in arbitrary units). At the critical lines $A$ and $B$ in Fig. 1(a) the spectral gap vanishes through the appearance of a pair of time-reversal symmetric band crossings with linear dispersion around zero-energy \cite{Malard20} in the BZ. When $\theta$ is perturbed along $A$ or $\phi$ is perturbed along $B$, the nodes move symmetrically through the one-dimensional BZ, similar to Weyl cones in a three-dimensional topological semimetal \cite{Murakami,Wan}, with the difference that here the nodes do not pairwise merge and annihilate at the center or at the boundaries of the BZ. While the nodes associated to $B$ are accidental, those of $A$ are enforced by the mirror symmetry present at $\phi=\pi/4$\, \cite{Malard20}. In Ref. \onlinecite{Malard20} it was also found that the TQPTs across $A$ or across $B$ are signaled through the appearance of a cusp, i.e. a nonanalyticity, in the second derivative of the ground state energy, and that the gap closes linearly with respect to the control parameter in these cases. At the multicritical points, the nonanalyticity of the ground state energy is pushed to fourth order, while the closing of the gap with respect to the control parameter becomes parabolic.

\section{Curvature function}

For the purpose of analyzing quantum criticality near the TQPTs and near the peculiar multicritical point, we turn to the method proposed in Ref.~\onlinecite{Chen3} and \onlinecite{Chen4} to investigate the curvature function. Generally, the curvature function $F({\bf M},k)$ is defined as the function of momentum and Hamiltonian parameters whose integration over the BZ yields the prescribed topological invariant \cite{Chen4}. From Eq. (\ref{Walternative}), it follows that
\begin{equation} \label{defF}
F({\bf M},k)\,=\,-\frac{1}{2\pi i}\frac{\partial_{k}\,\text{det}[Q({\bf M},k)]}{\text{det}[Q({\bf M},k)]}.
\end{equation}
A direct calculation shows that
\begin{eqnarray} \label{detQ}
\nonumber \text{det}[Q({\bf M},k)]&=&a\cos(2k)+b\cos(k)+c \nonumber \\
&+&i\left(-a\sin(2k)-b\sin(k)\right),
\end{eqnarray}
implying that
\begin{equation} \label{F}
F({\bf M},k)\!=\!\frac{1}{2\pi}\frac{2a^2\!+\!b^2\!+\!b(3a\!+\!c)\cos(k)\!+\!2ac\cos(2k)}{a^2\!+\!b^2\!+\!c^2\!+\!2b(a\!+\!c)\cos(k)\!+\!2ac\cos(2k)},
\end{equation}
with real parameters $a$, $b$ and $c$ given in terms of the amplitudes in Eqs. (\ref{alphabeta}) as
\begin{eqnarray} \label{parameters}
\nonumber a\,&=&\,(|\alpha_{2}^{+}\alpha_{4}^{+}|)^2+(\beta_{2}\beta_{4})^2-(|\alpha_{2}^{+}|\beta_{4})^2-(|\alpha_{4}^{+}|\beta_{2})^2, \\
\nonumber c\,&=&\,(|\alpha_{1}^{+}\alpha_{3}^{+}|)^2+(\beta_{1}\beta_{3})^2-(|\alpha_{1}^{+}|\beta_{3})^2-(|\alpha_{3}^{+}|\beta_{1})^2, \\
\nonumber b\,&=&\,-2\Re(\alpha_{1}^{+}\alpha_{2}^{+})\beta_{3}\beta_{4}-2\Re(\alpha_{1}^{+}\alpha_{3}^{-})\beta_{2}\beta_{4}-\\
\nonumber & &-2\Re(\alpha_{1}^{+}\alpha_{4}^{+})\beta_{2}\beta_{3}-2\Re(\alpha_{2}^{+}\alpha_{3}^{+})\beta_{1}\beta_{4}-\\
\nonumber & & -2\Re(\alpha_{2}^{+}\alpha_{4}^{-})\beta_{1}\beta_{3}-2\Re(\alpha_{3}^{+}\alpha_{4}^{+})\beta_{1}\beta_{2}-\\
& & -2\Re(\alpha_{1}^{+}\alpha_{2}^{+}\alpha_{3}^{+}\alpha_{4}^{+})-2\beta_{1}\beta_{2}\beta_{3}\beta_{4}.
\end{eqnarray}

Fig. 1(b) shows plots of $F$ as a function of $k$ for fixed $\gamma_{\text{eff}}=2.5$ and $\theta=0.1\pi$, with $\phi$ varying along the green path shown in Fig. 1(a) (parametrizing a vertical TQPT across critical line $A$), as well as for $\theta$ and $\phi$ varying along the magenta path shown in Fig. 1(a) (parametrizing a path going through the left multicritical point). We find that $F$ displays a double-peak structure in momentum space, and that the peaks are located away from the high-symmetry poins (HSPs) $k_{0}=0$ or $\pi$. The configuration with two down (up) peaks corresponds to the topologically trivial (nontrivial) gapped phase where the momentum integration of $F$ results in $W=0$ ($W=2$), in agreement with Fig. 1(a). As the system approaches the transition, the peaks gradually move to the two points in the BZ where the gap vanishes, while gradually narrowing and reaching their extreme points as $\phi\rightarrow\phi_{c}$. This asymptotic behavior of $F$, with a peak around each gap-closing point, is in agreement with the prediction of Ref. [\onlinecite{Chen4}] for linear band crossings. However, while our model yields a curvature function with two unpinned peaks - the peaks move symmetrically in the BZ as a Hamiltonian parameter is varied, following the movement of the associated band crossings in the spectrum (c.f. Sec. III) - the curvature function of the linear Dirac model discussed in Ref. [\onlinecite{Chen4}] contains only one single peak always located at the HSP. Interestingly, despite the difference in the characters of the TQPT (second-order nonanalyticity) and the multicritical point (fourth-order nonanalyticity), the critical behavior of the curvature function is found to be very similar in that the two peaks of $F$ diverge as either type of critical point is approached. The only difference is that the two peaks do not flip as the system passes through the multicritical point, a consequence of the unchangedness of the topological invariant. Since the critical behavior is extracted solely from the narrowing of the peaks with no consideration of the flipping \cite{Chen3,Chen4}, this suggests that the TQPTs and the multicritical point have a similar critical behavior.

To quantify the scaling properties of the two cases above, we note that each peak of $F$ can be fitted by a Lorentzian function of the form
\begin{equation} \label{Ffit}
F_{\text{fit}}(k)\,=\,\frac{1}{2\pi}+\frac{h}{1+\xi^{2}(k-k_{+})^{2}},
\end{equation}
where $k_{+}$ is the location of the peak along the positive $k$-axis, $h$ is the height of the peak and $\xi^{-1}$ its width. We now define the critical exponent $\gamma$ for $h$ and $\nu$ for $\xi$ by
\begin{eqnarray}
\label{hinv} h^{-1}(\lambda)\,&=&\,\pm C\,|\lambda-\lambda_{c}|^{\gamma},\\
\label{xiinv} \xi^{-1}(\lambda)\,&=&\,D\,|\lambda-\lambda_{c}|^{\nu},
\end{eqnarray}
where $C$ and $D$ are (positive) non-universal coefficients, and $\lambda$ parameterizes a generic path in the $\theta\times\phi$ parameter space across a single critical line or a multicritical point, with $\lambda_c$ defining the intersection of this path with the critical line or the multicritical point. In Eq. (\ref{hinv}), the plus (minus) sign applies for an up (down) peak. By fitting Eq. (\ref{Ffit}) to Eq. (\ref{F}) we extract the values of $h^{-1}$ and $\xi^{-1}$ (as well as of $k_{+}$) for selected paths in the phase diagram of Fig. 1(a). Fig. 2 shows data points collected for $h^{-1}$ (top) and for $\xi^{-1}$ (bottom) for paths through (a) critical line $A$, (b) critical line $B$, and (c) the left multicritical point in Fig. 1(a). Now fitting Eq. (\ref{hinv}) [(\ref{xiinv})] to the data collected for $h^{-1}$ [$\xi^{-1}$] we get the critical exponent $\gamma$ [$\nu$], as well as the coefficient $C$ [$D$]. These fits correspond to the continuous curves in Fig. 2, with the corresponding fitting parameters listed in the chart below the plots.
\begin{figure*}[htpb]
\includegraphics[width=18cm]{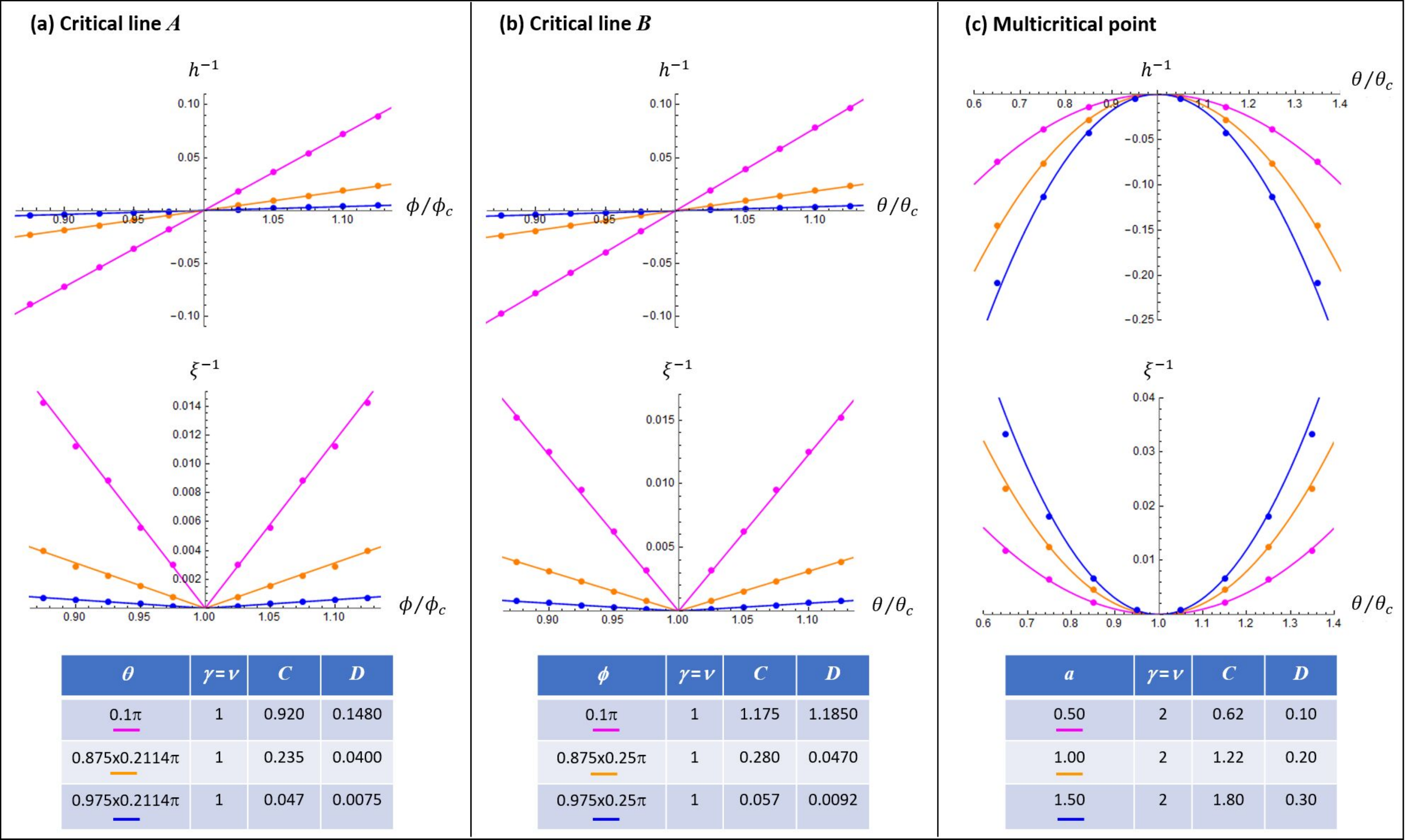}
\caption{(Color online) Data points and fitting of inverse height $h^{-1}$ (top) and width $\xi^{-1}$ (bottom) for (a) TQPTs through the critical line $A$ in Fig. 1(a), with fixed $\theta$ and $\lambda=\phi$ in Eqs. (\ref{hinv})-(\ref{xiinv}); (b) TQPTs through the critical line $B$ in Fig. 1(a), with fixed $\phi$ and $\lambda=\theta$ in Eqs. (\ref{hinv})-(\ref{xiinv}); (c) transitions parametrized by $(\phi-0.25\pi)=a(\theta-0.2114\pi)$, i.e. through the left multicritical point in Fig. 1(a), with $\lambda=\theta$ as the driving parameter in Eqs. (\ref{hinv})-(\ref{xiinv}). The results indicate that the critical exponents defined with respect to the control parameter $\lambda=\left\{\theta,\phi\right\}$ depend on the type of transition: $\gamma=\nu=1$ for the TQPTs and $\gamma=\nu=2$ for the multicritical point. However, when the exponents are defined with respect to the spectral gap $\Delta$, one obtains a universal scaling behavior $h^{-1}\propto\xi^{-1}\propto\Delta$ valid for both types of transitions.}
\end{figure*}

Figs. 2 (a)-(b) show that the critical exponents $\gamma=\nu=1$ for the TQPTs across a single critical line, even for transitions passing very close to the multicritical point (orange and blue). Transitions connecting phases with the same $W$ have $\gamma=\nu=2$, regardless of the slope $a$ of the path cutting through the multicritical point, as indicated by Fig. 2 (c). The scaling law $\gamma=D\nu$ derived in Ref. [\onlinecite{Chen4}] for a $D$-dimensional linear Dirac model is thus verified in our $D=1$ model for both the TQPTs (paths across either an $A$ {\em or} $B$ critical line) and through a multicritical point (where one region of a phase simply gets pinched off from another region of the same phase). We note that the scaling of $h^{-1}$ and of $\xi^{-1}$ become anomalous for TQPTs along paths which are not perpendicular to $A$ or to $B$ (not shown in Fig. 2), with different critical exponents $\gamma$ and $\nu$ at opposite sides of the transition. This happens because, along such ``tilted" TQPTs, the multicritical point affects the curvature function differently at opposite sides of the transition point, with comparatively sharper peaks along the segment of the transition which is closer to the multicritical point, thus disrupting the scaling.

A vanishing $\xi^{-1}$ near the transitions (c.f. Fig. 2) corresponds to a diverging length $\propto\xi$ in position space, the latter being connected to a vanishing energy scale - here given by the spectral gap $\Delta$ to the first excited level - according to $\Delta\backsim\xi^{-z}$, where $z$ is the dynamic critical exponent \cite{Sachdev}. This relation and Eq. (\ref{xiinv}) lead to the following expression:
\begin{equation} \label{Delta}
\Delta(\lambda)\,=\,J|\lambda-\lambda_{c}|^{z\nu},
\end{equation}
with $J$ a non-universal coefficient. In Ref. \onlinecite{Malard20} the behavior of $\Delta$ with the distance to a critical point was analyzed for various paths in the phase diagram of Fig. 1(a). It was found that $z\nu=1$ for TQPTs across a single critical line and $z\nu=2$ for the multicritical point. Combining this result with the present findings for $\nu$ and $\gamma$ (c.f. Fig. 2) yields a dynamic critical exponent $z=1$, and universal scaling relations
\begin{equation} \label{UniversalScalingWithGap}
h^{-1}\propto\xi^{-1}\propto\Delta,
\end{equation}
valid irrespective of the order of the transition, and of the path through the transition. This result shows that while transitions of different orders can be driven depending on the path taken in the Hamiltonian parameter space, there is still only one energy scale, namely the spectral gap, which underpins the universal scaling behavior. Moreover, the scaling law $\gamma=\nu$ that stems from the conservation of the topological invariant\cite{Chen3,Chen4} dictates that $h^{-1}$ and $\xi^{-1}$ must scale with $\Delta$ in the same manner. The distinction between normal criticality and multicriticality reveals itself only at the level of the individual scalings of $h$, $\xi$, and $\Delta$ with the control parameter $\theta$ or $\phi$.

\section{Skew-polarization and Wannier state correlation function}

In this section, we demonstrate that $\xi$ can be interpreted as a correlation length between bulk Wannier states. This interpretation follows from the generalization of the formula connecting the winding number $W$ of a two-band chiral-invariant Bloch Hamiltonian in $D=1$ to a ``skew-polarization" $\tilde{A}(k)$ \cite{Mondragon},
\begin{equation} \label{skew}
W = \frac{1}{\pi} \int_{-\pi}^{\pi} \tilde{A}(k) dk, \ \ \tilde{A}(k) = i  \sum_{\alpha} \langle u_{\alpha}(k)|{\cal S}\partial_k|u_{\alpha}(k)\rangle,
\end{equation}
where ${\cal S}$ is the chiral symmetry operator and $|u_{\alpha}(k)\rangle$ the Bloch function for band $\alpha$, with $\alpha$ summed over all occupied bands. This expression is similar to that from the theory of polarization\cite{Resta,Vanderbilt,VanderbiltBook} which relates the electronic polarization to the Wannier centers of the occupied bands. In the present case, the skew-polarization carries information about the distribution of the localized Wannier functions of the system. In the following we substantiate this expectation by generalizing Eq. (\ref{skew}) to \emph{any} multiband chiral-invariant Bloch Hamiltonian, in particular our $2r$-band model in Eqs. (\ref{Hk})-(\ref{alphabeta}) with even $r$.

We start from Eq. (\ref{defF}) and note that $Q$ in Eqs. (\ref{Qk})-(\ref{A}) is an invertible matrix which allow us to use the trace-determinant formula and rewrite the curvature function as
\begin{equation} \label{defF2}
F(k)\,=\,-\frac{1}{2\pi i}\text{Tr}[Q^{-1}(k)\,\partial_{k}Q(k)].
\end{equation}
where, to ease notation, we have suppressed the dependence on the parameter vector ${\bf M}$. Next, let us look at the matrix product ${\cal S}\,{\cal H}^{-1}\,\partial_{k}{\cal H}$, where ${\cal H}$ is the Bloch Hamiltonian of Eq. (\ref{Hk}) and ${\cal S}=\sigma_{z}\otimes1\!\!1_{r\times r}$ is the associated chiral symmetry matrix. It follows that
\begin{equation}
\nonumber {\cal S}{\cal H}^{-1}(k)\partial_{k}{\cal H}(k)\!=\!\begin{bmatrix}
    Q^{\dag^{-1}}(k)\partial_{k}Q^{\dag}(k) & 0 \\
    0 & -Q^{-1}(k)\partial_{k}Q(k) \\
\end{bmatrix}\!,
\end{equation}
and thus
\begin{eqnarray}
\nonumber & &\text{Tr}[{\cal S}{\cal H}^{-1}(k)\partial_{k}{\cal H}(k)] \\
\nonumber &=&(\text{Tr}[Q^{-1}(k)\partial_{k}Q(k)])^{\ast}\!-\!\text{Tr}[Q^{-1}(k)\partial_{k}Q(k)],
\end{eqnarray}
where we have used that $\text{Tr}[A^{\dag}B^{\dag}]=(\text{Tr}[AB])^{\ast}$. Evoking now Eq. (\ref{defF2}) and the fact that $F$ is a real function by construction, we arrive at
\begin{equation}
\nonumber F(k)\,=\,\frac{1}{4\pi i}\text{Tr}[{\cal S}\,{\cal H}^{-1}(k)\,\partial_{k}{\cal H}(k)]
\end{equation}
or, equivalently,
\begin{equation}\label{defF3}
F(k)\,=\,\frac{1}{4\pi i}\sum_{n,x}\langle u_{n,x}(k)|{\cal S}\,{\cal H}^{-1}(k)\,\partial_{k}{\cal H}(k)|u_{n,x}(k) \rangle,
\end{equation}
where $|u_{n,x}(k)\rangle$ is one of the $2r$ Bloch states at a given momentum $k$, with $n=1,..,r$ and $x=\pm$ where $+$ $(-)$ labels empty (filled) states. By labelling the states in this way we are taking into account that the band structure is half-filled (c.f. Sec. II).

To compute the expectation value in Eq. (\ref{defF3}) we start with
\begin{equation} \label{eigen}
\nonumber\partial_{k}{\cal H}|u\rangle = \partial_{k}\varepsilon |u\rangle + \varepsilon \partial_{k}|u\rangle - {\cal H}\,\partial_{k}|u\rangle,
\end{equation}
where $\epsilon$ is an energy eigenvalue, and, for ease of notation, we have omitted the indices $n$ and $x$, as well as the momentum $k$. We thus get
\begin{eqnarray}
\nonumber & & \langle u|{\cal S}{\cal H}^{-1}\partial_{k}{\cal H}|u\rangle \\
\nonumber &\!=\!\!&\partial_{k}\varepsilon\langle u|{\cal S}{\cal H}^{-1}|u\rangle\!+\!\varepsilon\langle u|{\cal S}{\cal H}^{-1}\partial_{k}|u\rangle\!-\!\langle u|{\cal S}\,\partial_{k}|u\rangle\!.
\end{eqnarray}
The first term on the right hand side of this expression vanishes since $\langle u|{\cal S}{\cal H}^{-1}|u\rangle=\varepsilon^{-1}\langle u|{\cal S}|u\rangle=0$, as follows from the orthogonality of ${\cal S}|u\rangle$ and $|u\rangle$. Moreover, using the chiral symmetry property ${\cal S}{\cal H}^{-1}=-{\cal H}^{-1}{\cal S}$, we can write the second term,
$\varepsilon\langle u|{\cal S}{\cal H}^{-1}\partial_{k}|u\rangle$, on the right-hand side as $-\langle u|{\cal S}\partial_{k}|u\rangle$. It follows that
\begin{equation}
\nonumber \langle u|{\cal S}{\cal H}^{-1}\partial_{k}{\cal H}|u\rangle\,=\,-2\langle u|{\cal S}\,\partial_{k}|u\rangle,
\end{equation}
and thus, from Eq. (\ref{defF3}), that
\begin{equation}
\nonumber F(k)\,=\,-\frac{1}{2\pi i}\sum_{n,x}\langle u_{n,x}(k)|{\cal S}\,\partial_{k}|u_{n,x}(k)\rangle.
\end{equation}
The summation over all states in the above expression can be restricted because the contribution from the empty states equals that of the filled ones. Indeed, since $|u_{+}\rangle={\cal S}|u_{-}\rangle$, we have that $\langle u_{+}|{\cal S}\,\partial_{k}|u_{+}\rangle=\langle u_{-}|{\cal S}^{\dag}{\cal S}\,\partial_{k}{\cal S}|u_{-}\rangle=\langle u_{-}|\partial_{k}{\cal S}|u_{-}\rangle=\langle u_{-}|{\cal S}\partial_{k}|u_{-}\rangle$. With that we arrive at the final expression for the curvature function written in terms of the filled Bloch states:
\begin{equation} \label{defF4}
F(k)\,=\,-\frac{1}{\pi i}\sum_{n}\langle u_{n,-}(k)|{\cal S}\,\partial_{k}|u_{n,-}(k)\rangle.
\end{equation}

As mentioned above, the relation in Eq. (\ref{defF4}) has been previously proposed in the context of a two-band chiral-invariant model in $D=1$ in Ref. \onlinecite{Mondragon}, and later also in Ref. \onlinecite{Chen4}. In both cases, the derivation depends on the basis of the Hamiltonians. In contrast, our basis-independent formalism above relies only on two general assumptions: (i) the presence of chiral symmetry and (ii) that the topological invariant is a $Z$-number defined as the winding, on the complex plane, of $\text{det}[Q]$, with $Q$ the off-diagonal matrix appearing in the chiral-symmetric Bloch Hamiltonian ${\cal H}$.

Finally, a Fourier transform yields the position-space curvature function \cite{Chen3,Chen4}
\begin{eqnarray}\label{Fpositionspace}
\tilde{F}(R)\,&=&\,-\frac{M}{\pi}\sum_{n}\langle \phi_{n,-}(0)|{\cal S}\,{\hat r}|\phi_{n,-}(R)\rangle
\nonumber \\
&=&\,-\frac{M}{\pi}\sum_{n}\int dr\,{\cal S}\,r\,W_{n}^{\ast}(r)W_{n}(r-R).
\end{eqnarray}
where $R$ and $0$ are the home cells of the Wannier states, ${\hat r}$ is the position operator, $M$ is the number of unit cells, and $|\phi_{n,-}\rangle$ are the filled bulk Wannier states whose Wannier functions are $\langle r|\phi_{n,-}\rangle=W_{n}(r-R)$. Eq. (\ref{Fpositionspace}) expresses $\tilde{F}(R)$ in terms of the overlap, weighted by the ${\cal S}\,{\hat r}$ operator, between Wannier states which are a distance $R$ apart.  We thus call $\tilde{F}(R)$ by \emph{Wannier state skew correlation function}. Analogous to the correlation functions in the usual Landau order-parameter paradigm for second-order phase transitions, the correlation function $\tilde{F}(R)$ measures how much the Wannier state at position $R$ remembers its configuration at the origin. This remembrance decays with a correlation length $\xi$ which diverges at the critical point, implying scale invariance at that point, as we demonstrate below.

We further make use of the Lorentzian fitting in Eq. (\ref{Ffit}) to write
\begin{equation}
\nonumber F(k)\,=\,F^{-}(k)\Theta(-k)+F^{+}(k)\Theta(k),
\end{equation}
where $F^{\pm}=F_{\text{fit}}$, with $+$\, ($-$) for a peak at $k_{+} (k_{-})$ in the expression for $F_{\text{fit}}$; $\Theta(k)$ is the Heaviside step function. Fourier transforming the above expression yields
\begin{equation}\label{Fpositionspace2}
\tilde{F}(R)\,=\,2\int_{0}^{\pi}F^{+}(k)\cos(kR)\,dk
\end{equation}
which can be easily evaluated numerically. Fig. 3 shows the result of $\tilde{F}(R)$ (a) for fixed $k_{+}$ and different values for $h$ and $\xi$ and (b) for fixed $h$ and $\xi$ and different values for $k_{+}$. The plots show that $\tilde{F}(R)$ is an oscillatory decaying function of $R$, a consequence of the double-peak structure of the momentum-space curvature (c.f. Fig. 1(b)). This is to be compared with the analogous result for a linear two-band Dirac model where a single-peaked momentum-space curvature yields a monotonically decaying $\tilde{F}(R)$ \cite{Chen3,Chen4}. Here the presence of two well-separated peaks in momentum-space generates periodic ``revivals" of the correlation between Wannier states throughout the chain.
\begin{figure}[t]
\begin{center}
\includegraphics[clip=true,width=0.7\columnwidth]{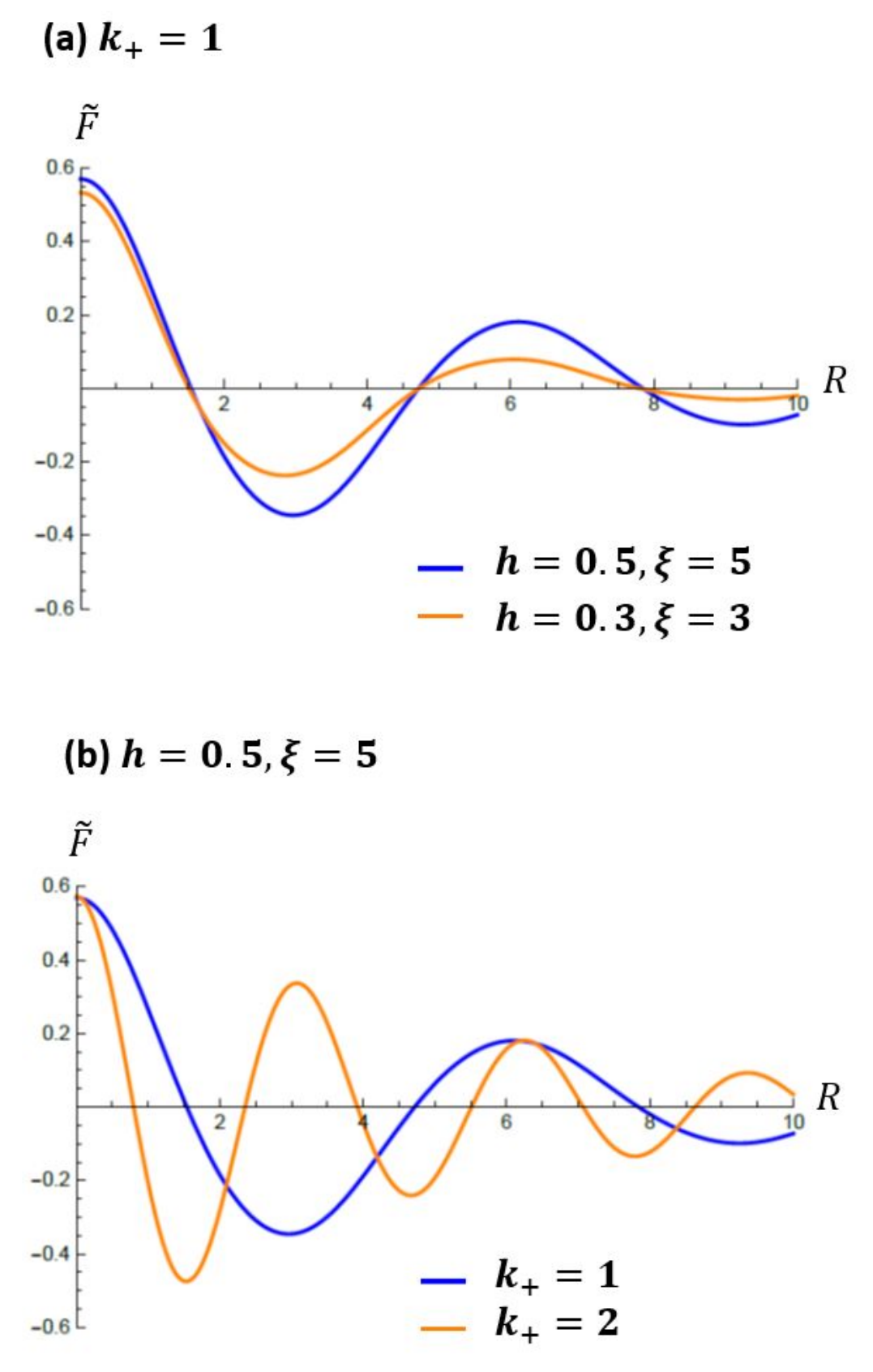}
\caption{(Color online) Wannier states skew correlation function $\tilde{F}$ as a function of the distance $R$ between the Wannier states for (a) fixed $k_{+}$ and different values for $h$ and $\xi$ and (b) fixed $h$ and $\xi$ and different values for $k_{+}$. $\tilde{F}$ is an oscillatory decaying function of $R$.}
\end{center}
\end{figure}

As seen in Fig. 3(a), the larger $\xi$, i.e. the narrower the peaks in momentum space, the longer the decay length in position space. Fig. 3(b) shows that the smaller the value of $k_{+}$, i.e. the closer the peaks are in momentum space, the longer the wave-length of the oscillation of $\tilde{F}(R)$. From Fig. 3(a) it becomes clear that $\xi$ plays the role of the correlation length of the Wannier state skew correlation function. The divergence of $\xi$ at the critical point then signifies that proximity to criticality makes the Wannier states correlated across longer distances and correlated over an infinitely long range at the critical point. The difference from the notion of scale invariance in a usual symmetry-breaking continuous phase transition is that, instead of converging to a constant everywhere, $\tilde{F}(R)$ remains oscillating as the system approaches the critical point. This is because the gap-closing points, and hence also the two peaks of the curvature $F(k)$, are forbidden by the underlying symmetries from merging at $k=0$ or at $k=\pm\pi$ \cite{Malard20}. As a result, the distance $2k_{+}$ between the peaks does not vanish as the critical point is approached, yielding a finite wave length for $\tilde{F}(R)$ arbitrarily close to criticality.

\section{Curvature renormalization group approach}

Although the critical behavior of our system is such that the peaks of $F$ occur at unpinned momenta away from the HSPs, as shown in Fig. 1(b), the curvature function at HSP $k_{0}=0$ and $k=\pm\pi$ in fact also senses the critical behavior, for $F(k)$ at different momenta are not independent: they sum up to be the topological invariant $W$ according to Eq.~(\ref{Walternative}) and (\ref{defF}). In this section, we show that the critical behavior at the HSPs, using $k_{0}=0$ as an example, allows the CRG approach \cite{Chen1,Chen2} to capture the phase diagram of Fig. 1(a) without explicitly performing the integration of Eq. (\ref{Walternative}). As we shall see, the fact that the bulk gap closes at non-HSPs $k_{\pm}$ renders the RG flow distinct from the usual Dirac models that have gap closing at a HSP.

The CRG approach is based on the iterative mapping ${\bf M}\rightarrow{\bf M}'$ that satisfies
\begin{equation}
F({\bf M},k_{0}+\delta k)=F({\bf M}',k_{0}),
\end{equation}
where $\delta k$ is a small deviation away from the HSP $k_{0}$. Expanding and keeping terms up to leading order yields the generic RG equation
\begin{equation}
\frac{dM_{i}}{d\ell}=\frac{M_{i}^{\prime}-M_{i}}{\delta k^{2}}=\frac{1}{2}\frac{\partial_{k}^{2}F({\bf M},k)|_{k=k_{0}}}{\partial_{M_{i}}F({\bf M},k_{0})},
\label{RG_eq_derivative}
\end{equation}
where $M_{i}=\left\{\theta,\phi\right\}$ is a component of the ${\bf M}$ vector, which may be evaluated numerically by
\begin{equation}
\frac{dM_{i}}{d\ell}=\frac{F({\bf M},k_{0}+\Delta k)-F({\bf M},k_{0})}{F({\bf M}+\Delta M_{i}{\bf\hat{M}}_{i},k_{0})-F({\bf M},k_{0})},
\end{equation}
where $\Delta k$ is a small (finite) deviation from the HSP in momentum space, and $\Delta M_{i}$ is a small interval in the parameter space along the ${\hat{\bf M}}_{i}$ direction. This numerical interpretation is a great advantage over the integration in Eq. (\ref{Walternative}), since for a point ${\bf M}$, one is only required to calculate the curvature function at three points $F({\bf M},k_{0}+\Delta k)$, $F({\bf M},k_{0})$ and $F({\bf M}+\Delta M_{i}{\hat{\bf M}}_{i},k_{0})$ to obtain the RG flow along the ${\hat{\bf M}}_{i}$ direction. The CRG is, therefore, a powerful tool to capture TQPTs in a multi-dimensional parameter space, as has been demonstrated for Floquet systems and interacting systems\cite{Molignini18,Chen18_weakly_interacting,Kourtis17,Molignini19,Panahiyan20}.
\begin{figure}[t]
\begin{center}
\includegraphics[clip=true,width=0.99\columnwidth]{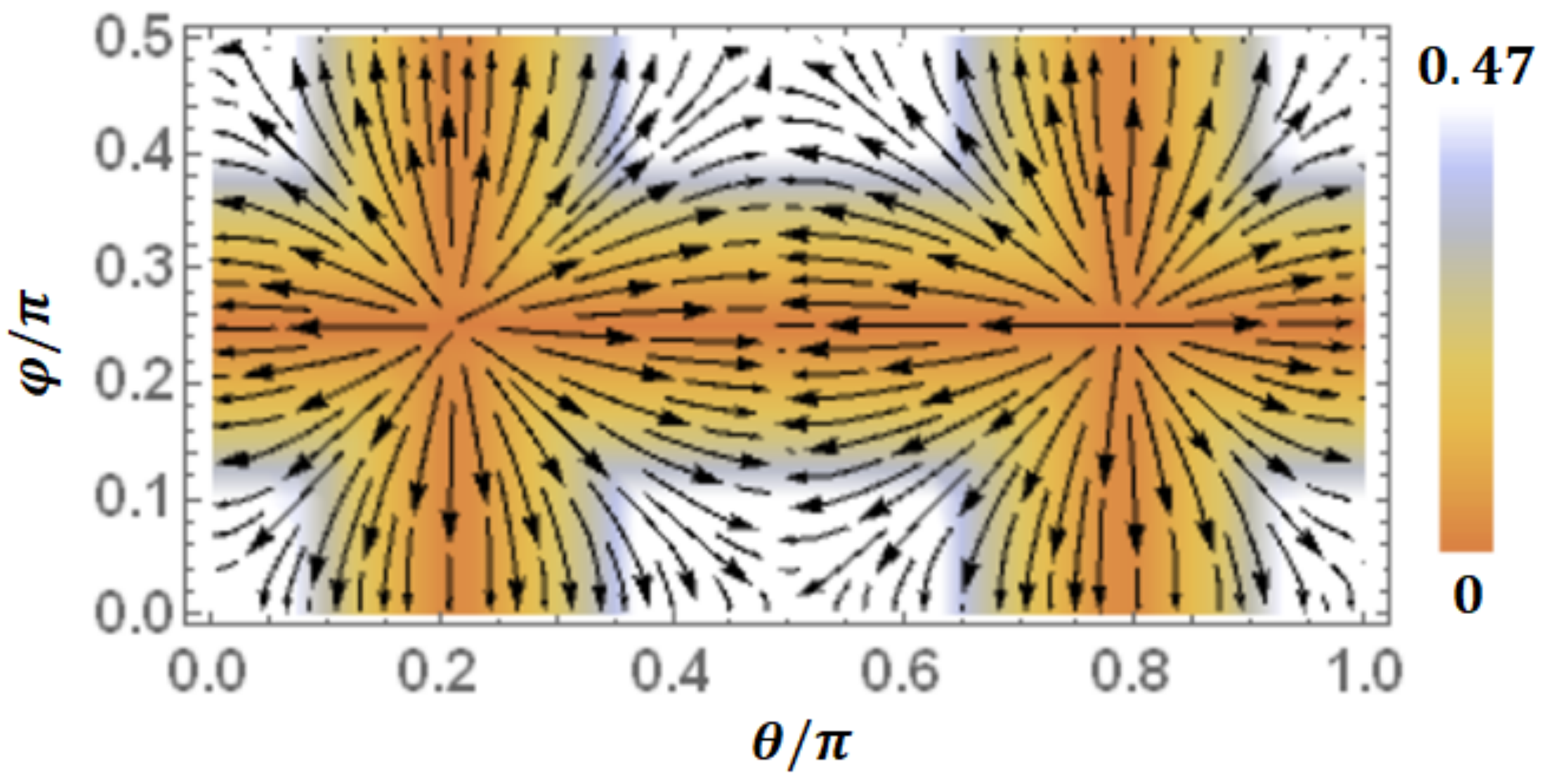}
\caption{(Color online) RG flow in the reduced parameter space ${\bf M}=(\theta,\phi)$,\emph{} with $\gamma_{\text{eff}} = 2.5$ (in arbitrary units), using a CRG approach. Arrows indicate the directions of the RG flow $d{\bf M}/d\ell=(d\theta/d\ell,d\phi/d\ell)$, and the color code indicate the magnitude ${\rm Min}(|d\theta/d\ell|,|d\phi/d\ell|)$. The orange lines where the magnitude goes to zero and from which the RG flows away coincide with the phase boundaries of the phase diagram of Fig. 1(a).}
\label{fig:RG_flow_results}
\end{center}
\end{figure}

For the usual Dirac models described by an $n$-dimensional parameter space, the phase boundary is an $(n-1)$-dimensional surface from which $d{\bf M}/d\ell$ flows away and where the magnitude $|d{\bf M}/d\ell|\rightarrow\infty$, a feature inherited from the divergence and flipping of the single peak of the curvature function at the HSP. On the other hand, the RG flows in the present model display a very distinct feature. To make comparison with the phase diagram in Fig. 1(a), we fix $\gamma_{\text{eff}}=2.5$ and obtain the RG flows on the two-dimensional parameter space ${\bf M}=(\theta,\phi)$ shown in Fig. 4. A comparison with Fig. 1(a) then shows that the phase boundaries coincide with those lines in Fig. 4 from which $d{\bf M}/d\ell=(d\theta/d\ell,d\phi/d\ell)$ flows away, but along which the magnitude of either $|d\theta/d\ell|$ or $|d\phi/d\ell|$ vanishes, instead of diverges \cite{ }. In other words, the phase boundaries manifest as lines of unstable fixed points, with the multicritical point being located where both components of the flow vanishes, $d\theta/d\ell=d\phi/d\ell=0$. A closer investigation reveals that the vanishing of $|d\theta/d\ell|$ or $|d\phi/d\ell|$ originates from the vanishing of the numerator of Eq.~(\ref{RG_eq_derivative}). This is because when ${\bf M}$ crosses the phase boundary, $\partial_{k}^{2}F({\bf M},k)|_{k=k_{0}}$ changes from slightly positive if the peaks at $k_{\pm}$ point upward to slightly negative if the peaks point downward, indicating that the curvature $F({\bf M},k)$ at $k_{0}$ and at $k_{\pm}$ are not independent. Although a vanishing $\partial_{k}^{2}F({\bf M},k)|_{k=k_{0}}$ does not give an additional meaningful length scale, it does serve to unambiguously signal the phase boundaries, with the great advantage that one needs not worry about the exact locations of the moving double peaks. The feature that the phase boundaries manifest as lines of unstable fixed points bears a striking resemblance with that uncovered recently in the extended\cite{Abdulla20} and periodically driven\cite{Molignini18} Kitaev $p$-wave superconducting chain, where the flipping of the curvature function at non-HSPs also yields unstable fixed points.

\section{Summary and Outlook}

We have carried out a scaling analysis of a model of spin-orbit coupled electrons subject to a spatial modulation in one dimension. The corresponding class-CII Hamiltonian \cite{Schnyder08,Kitaev09,Ryu10,Chiu2016} belongs to a large class of Aubry-Andr\'e-Harper-type models \cite{AubryAndre,Harper}, and exhibits a checkerboard phase diagram with topologically trivial and nontrivial phases characterized by distinct values of a winding number topological invariant \cite{Malard20}. Crossing a single phase boundary entails a TQPT signaled by a cusp in the second derivative of the ground state energy, accompanied by a linear closing of the spectral gap with respect to the control parameter. Regions belonging to the same topological phase are pinched off from each other by multicritical points where the ground state energy becomes nonanalytic at its fourth derivative, and where the gap closes parabolically \cite{Malard20}.

Our scaling analysis is based on investigating the momentum space curvature function whose integration yields the topological invariant. The curvature function in our model exhibits two peaks away from the HSPs and which move in the BZ when a Hamiltonian parameter is varied on a surface parallel and close to a critical surface, mirroring the motion of the gap closing points on the critical surface. The asymptotic behavior of the peaks near the TQPTs and the multicritical points allows to extract the critical exponents $\gamma$ and $\nu$ for the height and the width of the peaks, respectively. When defined with respect to the control parameters, these exponents satisfy $\gamma=\nu=1$ for the TQPTs across a single critical line and $\gamma=\nu=2$ for the multicritical points. However, while the value of these critical exponents depend on the type of transition, the scaling of the curvature peak with respect to the gap was found to be universal, as described by Eq. (\ref{UniversalScalingWithGap}). Moreover, the scaling law $\gamma=\nu$ inherited from the conservation of topological invariant is found to be always satisfied, independent of the type of transition, and for any path across a transition. This result indicates that the conservation of the topological invariant imposes a strong constraint which dictates a universal scaling law between the height and the width of the curvature peak, provided the curvature function evolves continuously at both sides of the transition.

Although the critical behavior takes place at the two peaks away from HSPs, the curvature function at the HSPs can still sense the critical behavior due to the conservation of the topological invariant. Based on this fact, the CRG approach can be applied to efficiently delineate the entire phase diagram. The phase boundaries were found to be lines of unstable fixed points of the renormalization group flow, similarly to what has been recently uncovered for models of $p-$wave superconductors \cite{Abdulla20,Molignini18}.

In addition to the above results, we have derived a Wannier state skew correlation function which measures the overlap, weighted by the skew polarization operator ${\cal S}\,{\hat r}$, of Wannier states that are a certain distance apart. Our basis-independent derivation is solely based on the symmetry properties of class CII and the well-established concept of skew polarization, and thus applies to any class CII models with arbitrary number of bands. In the considered model, the double-peak structure of the curvature function in momentum space yields a decaying and oscillatory Wannier state skew correlation function in real space. The inverse of the width of the momentum-space peak plays the role of the correlation length between the Wannier states which diverges at the critical point. On the other hand, the wave length of the Wannier state skew correlation function remains finite arbitrarily close to the critical point, a finding which adds to the monotonically decaying behavior found for linear two-band Dirac models where the curvature function formalism was originally developed \cite{Chen4}. Our results indicate that within the context of TQPTs, the concepts of critical exponents, scaling laws, correlation functions, and correlation length are not limited to second-order phase transitions, but also applicable and useful to describe higher order or multicritical phase transitions.

\section{ACKNOWLEDGMENTS}
H. J. acknowledges support from the Swedish Research Council through Grant No. 621-2014-5972, and W. C. is financed by the productivity in research fellowship from CNPq.


\begin{thebibliography}{}

\bibitem{Schnyder08}
A. P. Schnyder, S. Ryu, A. Furusaki, and A. W. W. Ludwig, Phys. Rev. B {\bf 78}, 195125 (2008).

\bibitem{Kitaev09}
A. Kitaev, AIP Conf. Proc. {\bf 1134}, 22 (2009).

\bibitem{Ryu10}
S. Ryu, A. P Schnyder, A. Furusaki, and A. W W Ludwig, New J. Phys. {\bf 12}, 065010 (2010).

\bibitem{Chiu2016}
C.-K. Chiu, J. C. Y. Teo, A. P. Schnyder, and S. Ryu, 
Rev. Mod. Phys. {\bf 88}, 035005 (2016).

\bibitem{Ando}
Y. Ando and L. Fu, 
Annu. Rev. Cond. Matt. Phys. {\bf 6}, 361 (2015).

\bibitem{Continentino1}
M. A. Continentino, F. Deus, and H. Caldas, 
Phys. Lett. A {\bf 378}, 1561 (2014).

\bibitem{Continentino2}
M. A. Continentino, 
Physica B: Cond. Matter {\bf 505}, A1 (2017).

\bibitem{RoyGoswamiSau}
B. Roy, P. Goswami, and J. D. Sau, 
Phys. Rev. B {\bf 94}, 041101(R) (2016).

\bibitem{Chen1}
W. Chen, 
J. Phys.: Condens. Matter, {\bf 28}, 055601 (2016).

\bibitem{Chen2}
W. Chen, M. Sigrist, and A. P. Schnyder, 
J. Phys.: Condens. Matter {\bf 28}, 365501 (2016).

\bibitem{Chen3}
W. Chen, M. Legner, A. R\"uegg, and M. Sigrist, 
Phys. Rev. B {\bf 95}, 075116 (2017).

\bibitem{Griffith}
M. A. Griffith and M. A. Continentino, 
Phys. Rev. E {\bf 97}, 012107 (2018).

\bibitem{Niewenburg}
E. P. L. van Nieuwenburg, A. P. Schnyder, and W. Chen, 
Phys. Rev. B {\bf 97}, 155151 (2018).

\bibitem{Chen4}
W. Chen and A. P. Schnyder, 
New. J. Phys. {\bf 21}, 073003 (2019).

\bibitem{Rufo}
S. Rufo, N. Lopes, M. A. Continentino, and M. A. R. Griffith, Phys. Rev. B {\bf 100}, 195432 (2019).

\bibitem{Continentino3}
M. A. Continentino, S. Rufo, and G. M. Rufo, in {\em Strongly Coupled Field Theories for Condensed Matter and Quantum Information theory}, A. Ferraz, K.S. Gupta, G. W. Semenoff, and P. Sodano (Eds.) (Springer, 2020).

\bibitem{Chen19_book_chapter}
W. Chen and M. Sigrist, in {\it Advanced Topological Insulators} Ch.~7, H. Luo (ed.) (Wiley-Scrivener, 2019).

\bibitem{Molignini20}
P. Molignini, W. Chen, and R. Chitra, Phys. Rev. B {\bf 101}, 165106 (2020).

\bibitem{Sachdev}
S. Sachdev, {\em Quantum Phase Transitions}, 2nd edition (Cambridge University Press, 2011).

\bibitem{ContinentinoBook}
M. Continentino, {\em Quantum Scaling in Many-Body Systems: An Approach to Quantum Phase Transitions}, 2nd edition (Cambridge University Press, 2017).

\bibitem{ChaikinLubensky}
P. M. Chaikin and T. C. Lubensky, {\em Principles of Condensed Matter Physics} (Cambridge University Press, 1995).

\bibitem{FisherBook}
M. E. Fisher, in {\em Multicritical Phenomena}, R. Pynn and A. Skjeltorp (eds.) (Plenum Press, 1984).

\bibitem{CarrBook}
C. Castelnovo, S. Trebst, and M. Troyer, in {\em Understanding Quantum Phase Transitions}, L. Carr (ed.)  (CRC Press, 2010).

\bibitem{Wen}
X.-G. Wen, Rev. Mod. Phys. {\bf 89}, 41004 (2017).

\bibitem{Xu}
C. Xu and S. Sachdev, 
Phys. Rev. B {\bf 79}, 064405 (2009).

\bibitem{Tupitsyn}
I. S. Tupitsyn, A. Kitaev, N. V. Prokof'ev, and P. C. E. Stamp, 
Phys. Rev. B {\bf 82}, 085114 (2010).

\bibitem{Haldane}
F. D. M. Haldane, Phys. Rev. Lett. {\bf 61}, 2015 (1988).

\bibitem{Sticlet}
D. Sticlet, L. Seabra, F. Pollmann, and J. Cayssol, Phys. Rev. B {\bf 89}, 115430 (2014).

\bibitem{Wakatsuki}
R. Wakatsuki, M. Ezawa, Y. Tanaka, and N. Nagaosa, Phys. Rev. B {\bf 90}, 014505 (2014).

\bibitem{Malard20}
M. Malard. D. Brandao, P. E. de Brito, and H. Johannesson, Phys. Rev. Research {\bf 2}, 033246 (2020).








\bibitem{AubryAndre}
S. Aubry and G. Andr\'e, Ann. Isr. Phys. Soc. {\bf 3}, (1980).

\bibitem{Harper}
P. G. Harper, Proc. Phys. Soc. Sec. A {\bf 68}, 874 (1955).

\bibitem{Ganeshan}
S. Ganeshan, K. Sun, and S. Das Sarma, Phys. Rev. Lett. {\bf 110}, 180403 (2013).

\bibitem{Malard18}
M. Malard, P. E. de Brito, S. \"Ostlund, and H. Johannesson, Phys. Rev. B {\bf 98}, 165127 (2018).


\bibitem{Murakami}
S. Murakami, New. J. Phys. {\bf 9}, 356 (2007).

\bibitem{Wan}
X. Wan, A. M. Turner, A. Vishwanath, and S. Y. Savrasov, Phys. Rev. B {\bf 83}, 205101 (2011).

\bibitem{Mondragon}
I. Mondragon-Shem, T. L. Hughes, J. Song, and E. Prodan, Phys. Rev. Lett. {\bf 113}, 046802 (2014).

\bibitem{Resta}
R. Resta, Ferroelectrics {\bf 136}, 51 (1992).

\bibitem{Vanderbilt}
R.D. King-Smith and D. Vanderbilt, Physical Review B {\bf 47}, 1651 (1993).

\bibitem{VanderbiltBook}
D. Vanderbilt, {\em Berry Phases in Electronic Structure Theory} (Cambridge University Press, 2018).

\bibitem{Molignini18}
P. Molignini, W. Chen, and R. Chitra, Phys. Rev. B {\bf 98}, 125129 (2018).

\bibitem{Chen18_weakly_interacting}
W. Chen, Phys. Rev. B {\bf 97}, 115130 (2018).

\bibitem{Kourtis17}
S. Kourtis, T. Neupert, C. Mudry, M. Sigrist, and W. Chen, Phys. Rev. B {\bf 96}, 205117 (2017).

\bibitem{Molignini19}
P. Molignini, R. Chitra, and W. Chen, EPL {\bf 128}, 36001 (2019).

\bibitem{Panahiyan20}
S. Panahiyan, W. Chen, and S. Fritzsche, arXiv:2007.10669.

\bibitem{Abdulla20}
F. Abdulla, P. Mohan, and S. Rao, arXiv:2003.10190.



\end{thebibliography}
\end{document}